\documentclass[final,5p,times,twocolumn,authoryear]{elsarticle}
\biboptions{numbers,square,comma,sort&compress}

\usepackage{amssymb}
\usepackage{amsmath}
\usepackage{float}
\usepackage{tabularx}
\usepackage{booktabs}
\usepackage{makecell}
\usepackage{multirow}

\usepackage{ragged2e}
\newcolumntype{C}{>{\Centering\arraybackslash}X} 

\journal{Materials Today Physics}

\begin{document}

\begin{frontmatter}

\title{\textbf{Hydrogen sublattice in LaH$_{12}$: structural stability and possible ordering motifs}}

\author[1]{Khasan Akhmadiev}

\author[2]{Di Zhou}
\author[3]{Viktor V. Struzhkin}

\author[1]{Alexander G. Kvashnin\corref{cor1}}
\ead{A.Kvashnin@skoltech.ru}

\author[2]{Dmitrii V. Semenok\corref{cor1}}
\ead{dmitrii.semenok@hpstar.ac.cn}

\cortext[cor1]{Corresponding authors}

\address[1]{Skolkovo Institute of Science and Technology, Moscow, Russia}
\address[2]{Center for High Pressure Science \& Technology Advanced Research (HPSTAR), Beijing, China}
\address[3]{Center for High Pressure Science \& Technology Advanced Research (HPSTAR), Shanghai, China}

\begin{abstract}
Lanthanum polyhydrides, synthesized under extreme pressure, exhibit near-room-temperature superconductivity. Finding a precise arrangement of the hydrogen sublattice that is critical for understanding the strong electron-phonon coupling remains experimentally unresolved due to the small X-ray scattering form factor of hydrogen atoms. In this work, we attempt to computationally reconstruct hydrogen sublattices of the LaH$_{12}$ at 100–200 GPa combining unconstrained evolutionary searches and structure generation with fixed lanthanum sublattice. By implementing a similarity metric, we screen the computationally generated configurations against experimental powder X-ray diffraction data. We specifically identify the trigonal $R\bar{3}$-LaH$_{12}$ and pseudo-hexagonal $C222_1$-La$_2$H$_{24}$ phases, which demonstrate close agreement with the experimental X-ray diffraction data. First-principles calculations confirm the metallic and superconducting nature of these hydrides. Crucially, our lattice dynamics analysis using the stochastic self-consistent harmonic approximation (SSCHA) indicates that these structures are dynamically stable. We reveal the presence of molecular H$_2$ units and soft phonon modes in the trigonal LaH$_{12}$ that suppresses superconductivity ($T_c \ge 30$~K), while the hydrogen clathrate network in the $C222_1$ phase drives strong electron-phonon interactions, yielding the coupling parameters of $\lambda\approx 1.83$, $\omega_{\log} \approx 87.5$~meV and transition temperature of $T_c \approx 150$~K at 150 GPa.
\end{abstract}




\end{frontmatter}


\section{Introduction}\label{sec1}

Lanthanum superhydrides are among the highest-$T_c$ conventional superconductors known: thermodynamically stable $fcc$-LaH$_{10}$ demonstrates superconductivity near 250 K under megabar pressures \cite{bib1,bib34}. Syntheses in diamond anvil cells (DACs) at various pressures have yielded many different hydrogen- and deuterium-rich phases, specifically LaH$_{11}$ and LaH$_{12}$, which may exhibit superconducting properties $T_c$(onset) up to 280 K \cite{bib2}. A possible indication of this superconductivity was found in radio-frequency transmission and nuclear magnetic resonance ($^1$H NMR) relaxometry measurements \cite{bib2}. Although powder X-ray diffraction (XRD) resolves only heavy lanthanum sublattices, identifying a cubic Pm$\bar{3}$ phase at 115 GPa and a hexagonal phase refined as the $P6_3/mmc$ phase at 165 GPa \cite{bib2}, it does not locate hydrogen atoms due to their very small X-ray scattering form factor. Consequently, the precise hydrogen arrangement, which governs the electron-phonon coupling (EPC) responsible for the high-$T_c$ superconductivity, remains unresolved.

The determination of the precise hydrogen sublattice in high-pressure polyhydrides is a long-standing challenge. Although neutron diffraction on synthesized deuterides presents a potential experimental workaround, its application is constrained because deuteride compounds frequently exhibit altered thermodynamic stability and structural chemistry driven by zero-point energy (ZPE) effects \cite{bib17,bib21}. Furthermore, neutron diffraction requires larger sample volumes ($\sim1$ $mm^3$), which restricts the maximum achievable pressures in the experiments \cite{bib20}. As a result, the exact hydrogen arrangements remain highly difficult to understand for several essential phases. Notable examples include the widespread $cF80$-X$_4$H$_{25}$ phase (X = Y, Sm, Tb, ...) and other recently identified single-crystal structures (e.g. Y$_{13}$H$_{75}$) \cite{bib3}, as well as complex stoichiometries H$_6$S$_5$ \cite{bib5} and LaD$_{11}$ \cite{bib1}. Among these unresolved phases, the structure LaH$_{12}$ is special, because recent theoretical studies on highly symmetric analogues of this phase, namely $Pm\bar{3}$-ScH$_{12}$, MgH$_{12}$, and LuH$_{12}$ \cite{bib4}, predict high superconducting critical temperatures approaching 300 K. Thus, searching for the exact hydrogen arrangement in LaH$_{12}$ provides a crucial key to understanding the mechanisms driving near-room-temperature superconductivity in this structural family.

To overcome these experimental limitations, advanced computational methods of crystal structure search have become crucial \cite{bib22,bib23}. In this work, we describe an attempt to computationally reconstruct the hydrogen sublattice of LaH$_{12}$ and investigate its physical properties, thermodynamic and dynamic stability under high pressures. We employ a comprehensive search strategy utilizing both the USPEX evolutionary algorithm for unconstrained structural searches \cite{bib6} and the EnvXGen crystal structure generator to strictly place hydrogen atoms within a fixed experimental unit cell with one or two La atoms \cite{bib7}. Moreover, we introduce a quantitative structure similarity metric to filter the generated configurations. Through this approach, we identify motifs that match the experimental XRD profile while maintaining thermodynamic competitiveness on the convex hull, and dynamical stability within an anharmonic approximation \cite{bib16,bib24}.

\section{Computational methodology}\label{sec2}

\subsection{Crystal structure search}\label{subsec1}

To identify thermodynamically stable atomic configurations in La-H phases under megabar pressures, we employed a structure generation approach. First-principles calculations based on density functional theory (DFT) were carried out using the Vienna Ab Initio Simulation Package (VASP) \cite{bib9,bib10,bib11,bib12}. The projector-augmented wave (PAW) method was used to treat interactions of core-valence electrons \cite{bib13}, while the exchange-correlation energy was described by the generalized gradient approximation (GGA) with the functional of Perdew-Burke-Ernzerhof (PBE) \cite{bib14}. A plane-wave basis set was defined with a kinetic energy cutoff of 600 eV for all structural relaxations.

Global unconstrained structural searches were performed using the evolutionary algorithm USPEX \cite{bib6}, targeting the LaH$_{11}$, LaH$_{12}$ and LaH$_{13}$ stoichiometries, where Z (number of lanthanum atoms in the unitcell) is varied from 1 to 4. The initial population consisted of 120 structures, while subsequent population sizes comprised 100 structures. Generations were produced by applying evolutionary variation operators with the following distribution: 50\% heredity, 20\% random symmetric generation, 10\% random topological generation and 20\% soft mode mutation. The evolutionary algorithm was allowed to run for a maximum of 120 generations, employing an early stopping criterion if the best predicted structure remained unchanged for 36 consecutive generations. To accelerate the exploration of the potential energy surface, the evolutionary search was also coupled with a machine-learned interatomic potential, MatterSim \cite{bib8}. Structural relaxation and enthalpy evaluation were conducted in four sequential stages, where the Brillouin zone sampling was tightened from a $\Gamma$-centered Monkhorst-Pack (MP) \cite{bib15} $\mathbf{k}$-point grid spacing of 0.12 $\text{\AA}^{-1}$ to 0.06 $\text{\AA}^{-1}$.

To resolve the precise hydrogen sublattice within the established heavy-atom framework known from the XRD experiment, we utilized the EnvXGen algorithm \cite{bib7}. Candidate structures generated by EnvXGen were subjected to a two-step DFT structural relaxation protocol. In the initial stage, preliminary optimization was conducted with a constrained lanthanum sublattice to sample configurations with optimal hydrogen positions. The electronic convergence for the self-consistent field (SCF) cycles was set $10^{-4}$ eV, the structures were relaxed until the residual forces on each atom were less than $10^{-2}$ eV/\AA\ and a $\Gamma$-centered Monkhorst-Pack (MP) \cite{bib15} $\mathbf{k}$-point grid spacing of 0.20 $\text{\AA}^{-1}$ was utilized. In the subsequent stage, full unconstrained relaxation of atomic positions and lattice parameters was performed, establishing the electronic SCF convergence of $10^{-6}$ eV, a force threshold of $10^{-4}$ eV/$\text{\AA}$ and a $\mathbf{k}$-point spacing of 0.15 $\text{\AA}^{-1}$.

\subsection{Metric Evaluation}\label{subsec3}

Configurations generated by EnvXGen were screened against the experimentally known La sublattice using the ALIGNN descriptor \cite{bib27} as a preliminary filter, where a structural correspondence was assessed by cosine similarity:
\begin{equation}
    S = \frac{\mathbf{v}_{\text{exp}} \cdot \mathbf{v}_{\text{calc}}}{\|\mathbf{v}_{\text{exp}}\| \|\mathbf{v}_{\text{calc}}\|}
\end{equation}
where $\mathbf{v}_{\text{exp}}$ and $\mathbf{v}_{\text{calc}}$ represent the ALIGNN feature vectors for the experimental and calculated sublattices, respectively. In addition to assessing the generated structures, we implemented an integral metric for the direct comparison of diffractograms. A Gaussian broadening function was applied to both the experimental and theoretical discrete reflections to reconstruct intensity profiles. The broadened intensity profile $f(2\theta)$ for either datasets is generated as follows:
\begin{equation}
    f(2\theta) = \sum_{i} a_i \exp\left[ -\frac{(2\theta - 2\theta_i)^2}{2\sigma^2} \right]
\end{equation}
where $a_i$ and $2\theta_i$ are the intensity and position of the $i$-th Bragg peak, and $\sigma$ is the broadening parameter (set to 0.15$^\circ$). This procedure yields the continuous profiles $f_{\text{exp}}(2\theta)$ and $f_{\text{calc}}(2\theta)$. To account for slight pressure deviations that cause minor lattice volume variations, a linear angular shift $z_0$ was introduced to the calculated pattern:
\begin{equation}
    f_{calc}(2\theta, z_0) = \sum_{i} a_i \exp\left[ -\frac{(2\theta - 2\theta_i + z_0)^2}{2\sigma^2} \right]
\end{equation}
The optimal shift parameter $z_0$ was determined by minimizing the absolute area difference between the reconstructed experimental pattern and the shifted theoretical pattern using a gradient-free optimization algorithm:
\begin{equation}
    \mathcal{E} = \min_{z_0} \int_{0}^{2\theta_{\text{max}}} \big\vert f_{\text{exp}}(2\theta) - f_{\text{calc}}(2\theta, z_0) \big\vert \, d(2\theta)
\end{equation}
This metric enabled the isolation of high-symmetry configurations, which optimally reproduced the experimental XRD profiles.

\subsection{Electronic and thermodynamic properties}\label{subsec:electron_thermo_props}

To evaluate the dynamic stability of the obtained high-symmetry structures and construct the thermodynamic convex hull, we calculated harmonic phonon dispersions and phonon densities of states (DOS). These calculations were performed using the finite displacement method as implemented in the Phonopy code \cite{bib25,bib26}. The interatomic forces were computed using VASP with $2 \times 2 \times 2$ grid supercells, an electronic SCF convergence criterion of $10^{-8}$ eV and an atomic displacement amplitude of 0.01 \AA. Thermodynamic stability was assessed by calculating the formation enthalpies of the generated structures and constructing the convex hull phase diagram, explicitly incorporating the ZPE corrections estimated from the harmonic phonon spectra with removed imaginary modes.

To confirm the metallic nature of the obtained hydrides and quantify the orbital hybridization at the Fermi level, electronic band structures along high-symmetry $\mathbf{k}$-paths and projected DOS were computed. The electronic properties were calculated in the non-self-consistent regime using a pre-converged charge density with a convergence criterion of $10^{-6}$ eV. The tetrahedron method with Blöchl corrections was used on a finely resolved energy grid of 8000 points, coupled a $\Gamma$-centered Monkhorst-Pack (MP) \cite{bib15} $\mathbf{k}$-point spacing of 0.10 $\text{\AA}^{-1}$ and Gaussian smearing with a broadening parameter of 0.05 eV.

\subsection{Anharmonic lattice dynamics}\label{subsec:sscha}

To account for the large quantum fluctuations of hydrogen atoms and the anharmonicity of the Born-Oppenheimer potential energy surface, we employed the stochastic self-consistent harmonic approximation (SSCHA) \cite{bib18,bib19}. The SSCHA is a fully variational method in which the true free energy of the system $\mathcal{F}$ is minimized with respect to a trial harmonic density matrix $\tilde{\rho}_{\vec{\mathcal{R}}, \mathbf{\Phi}}$, which is parameterized by the centroid positions $\vec{\mathcal{R}}$ and auxiliary force constants $\mathbf{\Phi}$:
\begin{equation}\label{eq:sscha}
\mathcal{F}(\vec{\mathcal{R}}, \mathbf{\Phi}, \{\vec{a}_i\}) = F_{\mathbf{\Phi}} + \big\langle V(\mathbf{R}) - \mathcal{V}_{\vec{\mathcal{R}}, \mathbf{\Phi}}(\mathbf{R}) \big\rangle_{\tilde{\rho}_{\vec{\mathcal{R}}, \mathbf{\Phi}}(\mathbf{R})}
\end{equation}
\begin{equation}\label{eq:sscha_opt}
\mathcal{F}(\{\vec{a}_i\}) = \min_{\mathbf{\Phi},\vec{\mathcal{R}}} \mathcal{F}(\vec{\mathcal{R}}, \mathbf{\Phi}, \{\vec{a}_i\})
\end{equation}
where $F_{\mathbf{\Phi}}$ represents the analytical free energy of the trial harmonic Hamiltonian, $\mathcal{V}_{\vec{\mathcal{R}}, \mathbf{\Phi}}(\mathbf{R})$ is the trial harmonic potential, $V(\mathbf{R})$ is the true Born-Oppenheimer potential energy evaluated in instantaneous atomic configurations $\mathbf{R}$, and $\{\vec{a}_i\}$ are the three unit cell vectors. Within this framework, the spatial probability distribution of the ionic coordinates, given by the diagonal of the density matrix $\tilde{\rho}_{\vec{\mathcal{R}}, \mathbf{\Phi}}(\mathbf{R})$ in real space, is mathematically defined as a multidimensional Gaussian distribution. 

Due to intractable analytical integration of the exact anharmonic potential $V(\mathbf{R})$ for extended systems, the expectation value in Eq.~\ref{eq:sscha} is stochastically evaluated using Monte Carlo integration. A thermal ensemble of random atomic configurations $\mathbf{R}$ is generated according to the Gaussian probability distribution, and the true Born-Oppenheimer forces for each configuration are computed via DFT. An importance sampling reweighting technique is applied, allowing multiple minimization steps without recalculating DFT forces until the stochastic noise exceeds a critical effective sample size threshold. The parameters $\vec{\mathcal{R}}$ and $\mathbf{\Phi}$ are then iteratively updated using conjugate gradient minimization.

The DFT force calculations for the SSCHA populations were performed using the Quantum ESPRESSO (QE) package \cite{bib28} on $2\times2\times2$ grid supercells. The exchange correlation interaction was treated within the GGA and parametrized by PBE. PAW pseudopotentials were used setting the plane-wave kinetic energy of 50 Ry and charge density cutoffs of 500 Ry. The Brillouin zone of the supercell was sampled using a $6\times6\times6$ and $3\times3\times3$ $\mathbf{k}$-point mesh for both phases $R\bar{3}$-LaH$_{12}$ and $C222_1$-La$_2$H$_{24}$, respectively. SSCHA minimization was performed at 300 K, where each population consisted of 400 configurations. The number of populations needed to reach convergence depended on the structure and pressure: 12, 14 and 14 for $R\bar{3}$-LaH$_{12}$ at 100, 150 and 200 GPa, respectively; 17 for the constrained-lattice LaH$_{12}$ search at 150 GPa; and 18, 8 and 17 for $C222_1$-La$_2$H$_{24}$ at 100, 150 and 200 GPa, respectively. Final free energies and phonon spectra were taken from the last population.

\subsection{Electron-phonon coupling calculations}\label{subsec:epc}

The electron-phonon spectral function $\alpha^2F(\omega)$ was evaluated by combining anharmonic dynamical matrices obtained from the SSCHA with static electron-phonon matrix computed within density functional perturbation theory (DFPT). DFPT calculations were performed in QE on the primitive unitcell using a $\mathbf{k}$-mesh of 30$\times$30$\times$30 ($R\bar{3}$-LaH$_{12}$) and 20$\times$20$\times$20 ($C222_1$-La$_2$H$_{24}$) and the dynamical matrices obtained on the 2×2×2 mesh were Fourier-interpolated to 6×6×6 ($R\bar{3}$-LaH$_{12}$) and 4×4×4 ($C222_1$-La$_2$H$_{24}$) q-meshes for the electron-phonon calculations. We extracted phonon frequencies $\omega_{\mathbf{q}\nu}$ by diagonalizing the SSCHA dynamical matrices and calculated the spectral function:
\begin{equation}
\begin{split}
\alpha^2F(\omega) &= \frac{1}{N(E_F)} \sum_{nm,\nu} \int_{\text{BZ}} \frac{d^3k}{\Omega_{\text{BZ}}} \int_{\text{BZ}} \frac{d^3q}{\Omega_{\text{BZ}}} \\
&\quad \times |g_{nm}^{\nu}(\mathbf{k}, \mathbf{q})|^2 \delta(\epsilon_{n\mathbf{k}} - E_F) \delta(\epsilon_{m\mathbf{k}+\mathbf{q}} - E_F) \delta(\omega - \omega_{\mathbf{q}\nu})
\end{split}
\end{equation}
where $N(E_F)$ is the electronic DOS at the Fermi level $E_F$, $\epsilon$ are Kohn-Sham eigenvalues, $\{n,m\}$ are band indices, and $\mathbf{k}$ is the electron wavevector. A mode-resolved electron-phonon matrix $g_{nm}^{\nu}(\mathbf{k}, \mathbf{q})$ is determined by projecting the DFPT deformation potentials onto the anharmonic SSCHA polarization vectors:
\begin{equation}
g_{nm}^{\nu}(\mathbf{k}, \mathbf{q}) = \sum_{\kappa, \alpha} \frac{e_{\kappa\alpha}^{\nu}(\mathbf{q})}{\sqrt{2M_\kappa \omega_{\mathbf{q}\nu}}} \left\langle \psi_{m, \mathbf{k}+\mathbf{q}} \left\vert{} \frac{\partial V_{\text{KS}}}{\partial u_{\kappa\alpha}(\mathbf{q})} \right\vert{} \psi_{n, \mathbf{k}} \right\rangle
\end{equation}
where $M_\kappa$ is the mass of atom $\kappa$, $\mathbf{q}$ is the phonon wavevector, $e_{\kappa\alpha}^{\nu}(\mathbf{q})$ denotes the unit polarization vector of the phonon mode $\nu$ along the Cartesian direction $\alpha$, $\psi$ are Kohn-Sham wavefunctions and $\partial V_{\text{KS}} / \partial u_{\kappa\alpha}(\mathbf{q})$ is the linear variation of the electronic potential with respect to the atomic displacement. This mathematical projection incorporates quantum fluctuations into the final EPC evaluation.

The total EPC constant $\lambda$ and the logarithmic average phonon frequency $\omega_{\text{log}}$ were obtained from the integrated spectral function:
\begin{equation}
\lambda = 2 \int_{0}^{\infty} \frac{\alpha^2F(\omega)}{\omega} d\omega, \quad \omega_{\text{log}} = \exp \left( \frac{2}{\lambda} \int_{0}^{\infty} \frac{\alpha^2F(\omega)}{\omega} \ln\omega \, d\omega \right).
\end{equation}

To eliminate the artificial dependence of the EPC on the finite smearing width used for the Brillouin zone integration, the spectral function was renormalized using the tetrahedron method and scaled as:
\begin{equation}
\alpha^2F(\omega)_{\text{corr}} = \alpha^2F(\omega) \frac{N_{\text{tetra}}(E_F)}{N_{\text{smear}}(E_F)}
\end{equation}
where $N_{\text{tetra}}(E_F)$ and $N_{\text{smear}}(E_F)$ are the exact DOS evaluated via the tetrahedron method and the DOS obtained with the corresponding smearing scheme, respectively. Consequently, the total EPC constant scales linearly by the exact same factor to result $\lambda_{\text{corr}}$, while the logarithmic average frequency $\omega_{\text{log}}$ remains mathematically invariant under this uniform scaling. The superconducting transition temperature $T_c$ was then estimated using the Allen-Dynes equation \cite{bib35}:
\begin{equation}
T_c = f_1f_2\frac{\omega_{\text{log}}}{1.2} \exp \left[ -\frac{1.04(1+\lambda_{\text{corr}})}{\lambda_{\text{corr}} - \mu^*(1+0.62\lambda_{\text{corr}})} \right]
\end{equation}
where $\mu^*$ is the Coulomb pseudopotential representing the effective electron-electron repulsion, and $f_1f_2$ is the Allen-Dynes factor for strong-coupled superconductors \cite{bib35}. In our calculations using the Allen-Dynes formula, a standard value of $\mu^* = 0.10$ was adopted. The frequency-weighted spectral moment $A_n$ was evaluated to further quantify the EPC strength:
\begin{equation}
A_n = \int_0^{\omega_{\max}} \omega^{n} \alpha^2 F(\omega) d\omega
\end{equation}

As recently proposed for the screening of hydride superconductors \cite{bib33}, $A_n$ may serve as an additional descriptor of the electron-phonon interaction. The conventional EPC constant $\lambda$ scales as $\omega^{-1}$ and artificially diverges near lattice instabilities, whereas the $A_n$ (n$\ge$0) based formulation contains no inverse phonon frequency factors. Consequently, $A_n$ avoids numerical singularities and remains mathematically finite even in the presence of strong anharmonic phonon softening. This stability makes it a reliable metric for comparing the inherent EPC across structural phases.

\section{Results and Discussion}\label{sec3}

\begin{figure}[!htbp]
\centering
\includegraphics[width=0.98\columnwidth]{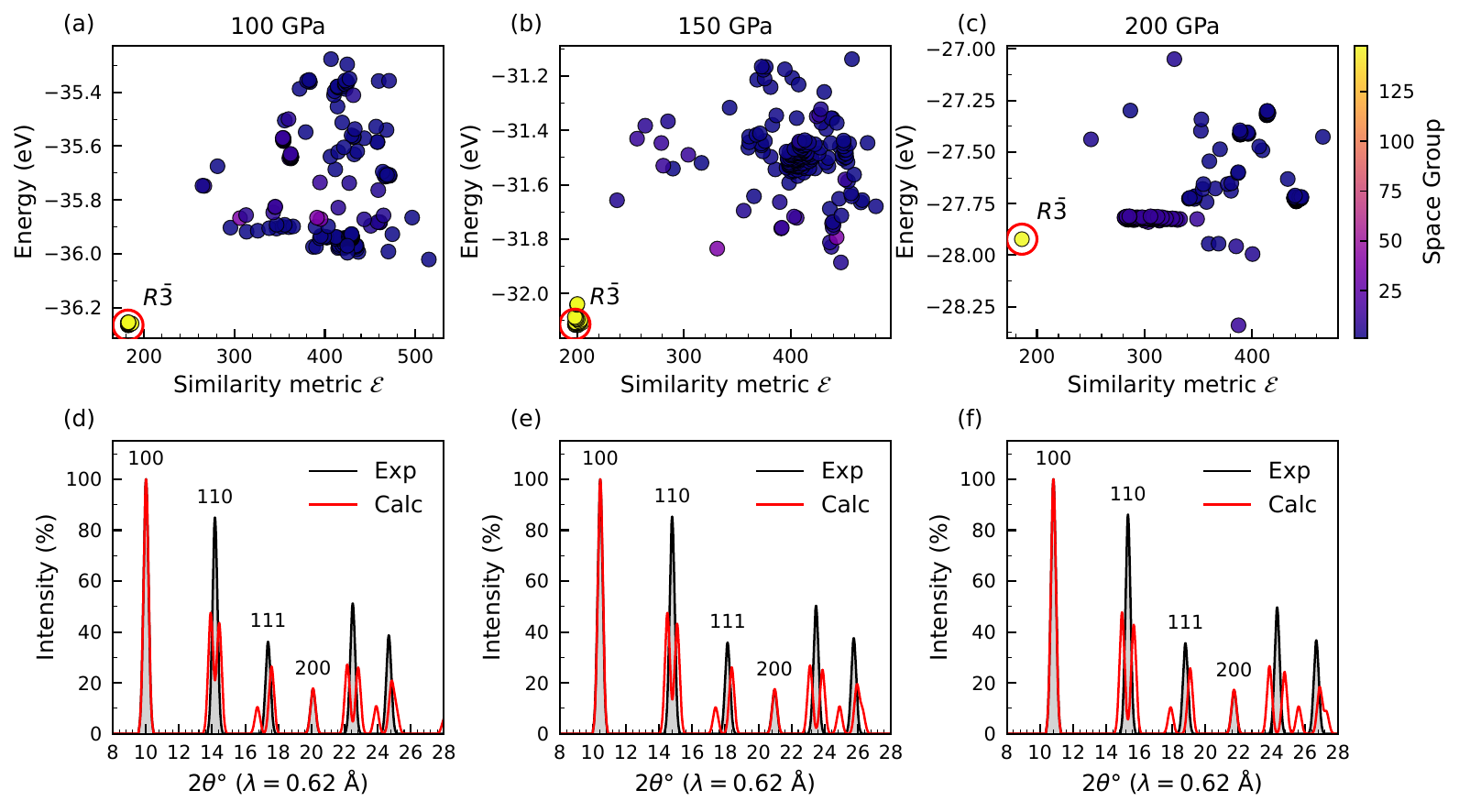}
\caption{Structural screening of the generated LaH$_{12}$ configurations with simple cubic La-sublattice at 100, 150 and 200 GPa. Color gradients indicating the respective space groups. The top panels (a-c) display the calculated energy versus the similarity metric. The optimal chosen phase highlighted by a red circle. The bottom panels (d-f) present the corresponding simulated XRD profiles of the optimal $R\bar{3}$ structures (red profile) compared with the experimental patterns of $Pm\bar{3}$ phase (black filled profiles). Distortion of the La-sublattice in the $R\bar{3}$-LaH$_{12}$ may be reduced by proper SSCHA relaxation.}\label{fig:integrals_sim_xrd}
\end{figure}

To explore the configuration space of the LaH$_{12}$ system, EnvXGen searches were performed under harmonic pressure at 100, 150 and 200 GPa (Figure~\ref{fig:integrals_sim_xrd}). Initially, La sublattice was fixed in the high-symmetry cubic cell shape ($Pm\bar{3}$) identified from the experimental diffraction data obtained at 115 GPa \cite{bib2}. The structure $Pm\bar{3}$-LaH$_{12}$, analogous to high-$T_c$ phases predicted for ScH$_{12}$ and LuH$_{12}$ \cite{bib4}, contains H$_2$ pseudo-molecules characterized by elongated interatomic distances ($d_{HH}>1.1$\AA). Although these quasi-molecular units represent a local potential energy minimum for Sc, Lu, and Mg hydrides in the harmonic and anharmonic approximations, our calculations reveal that this minimum vanishes for La polyhydride in both cases (see Figure S8 in the Supplementary Information). This instability may also point to the inherent limitations of the standard PBE functional, which does not accurately describe the configuration energies and properties of the molecular bond, leading to an underestimate of thermal and ZPE contributions \cite{bib29,bib30}.

Optimal candidates from our structural screening based on the tradeoff between enthalpy and the similarity metric cluster tightly, well separated from the rest of generated structures (Figure~\ref{fig:integrals_sim_xrd}a, b, c). Notably, repeating the 150 GPa structure search by comparing a constrained regime, where both La positions and cell shape were fixed, against a fully unconstrained structural relaxation alters the potential energy landscape but consistently converges on the identical $R\bar{3}$ motif as the best-scoring solution for Z = 1 (see Figures S1 b, c, e, f in the Supplementary Information). We consider this convergence and clear separation as the evidence that the structure $R\bar{3}$ is a reproducible hydrogen-ordering motif, intrinsic to the LaH$_{12}$ system and not a computational artifact. Specifically, these optimal candidates relax into the phase $R\bar{3}$-LaH$_{12}$, whose La-sublattice is slightly distorted phase $Pm\bar{3}$. In contrast to the structures mentioned ScH$_{12}$ and LuH$_{12}$ \cite{bib4}, where hydrogen atoms form pseudo-molecules that typically host strong electron-phonon coupling, the $R\bar{3}$ phase relaxes into a hydride with a true molecular hydrogen sublattice. Due to the presence of molecular units H$_2$ that localize electrons, the superconducting properties of the phase $R\bar{3}$ are expected to be lower.

To investigate the experimentally observed hexagonal LaH$_{12}$ \cite{bib2}, EnvXGen searches were performed within the La sublattice related to the space group  $P6_3/mmc$ using a minimal simulation cell with two La atoms (Z=2). These searches converge toward the pseudo-hexagonal structure $C222_1$-La$_2$H$_{24}$ as the most promising candidate (see Figures S1 (a, d) in the Supplementary Materials). In contrast to the cubic-derived molecular phase $R\bar{3}$, the $C222_1$-La$_2$H$_{24}$ possesses a dense distorted network of hydrogen bonds. Such an extended atomic-like hydrogen motif may lead to strong electron-phonon coupling, proposing a better candidate to explain the high $T_c$ possibly observed experimentally in lanthanum polyhydrides LaH$_{12}$.

Detailed analysis of the simulated XRD patterns reveals a distinct structural deviation of the relaxed phase $R\bar{3}$ from the initial $Pm\bar{3}$-LaH$_{12}$. The ordered network of H$_2$ molecules exerts an anisotropic local crystal field on the La atoms, forcing them to displace from their high-symmetry positions. This symmetry breaking leads to a splitting of the 110 XRD peak, as clearly visible in Figures~\ref{fig:integrals_sim_xrd}d, e, f. While this distortion lowers the potential energy, the subsequent SSCHA analysis will demonstrate that incorporating strong anharmonicity and quantum zero-point motion stabilizes both the distorted and undistorted structures. This indicates that quantum fluctuations are sufficient to dynamically average the orientations of the H$_2$ molecules, thereby restoring the highly symmetric lanthanum sublattice in this configuration.

\begin{figure}[!htbp]
\centering
\includegraphics[width=0.98\columnwidth]{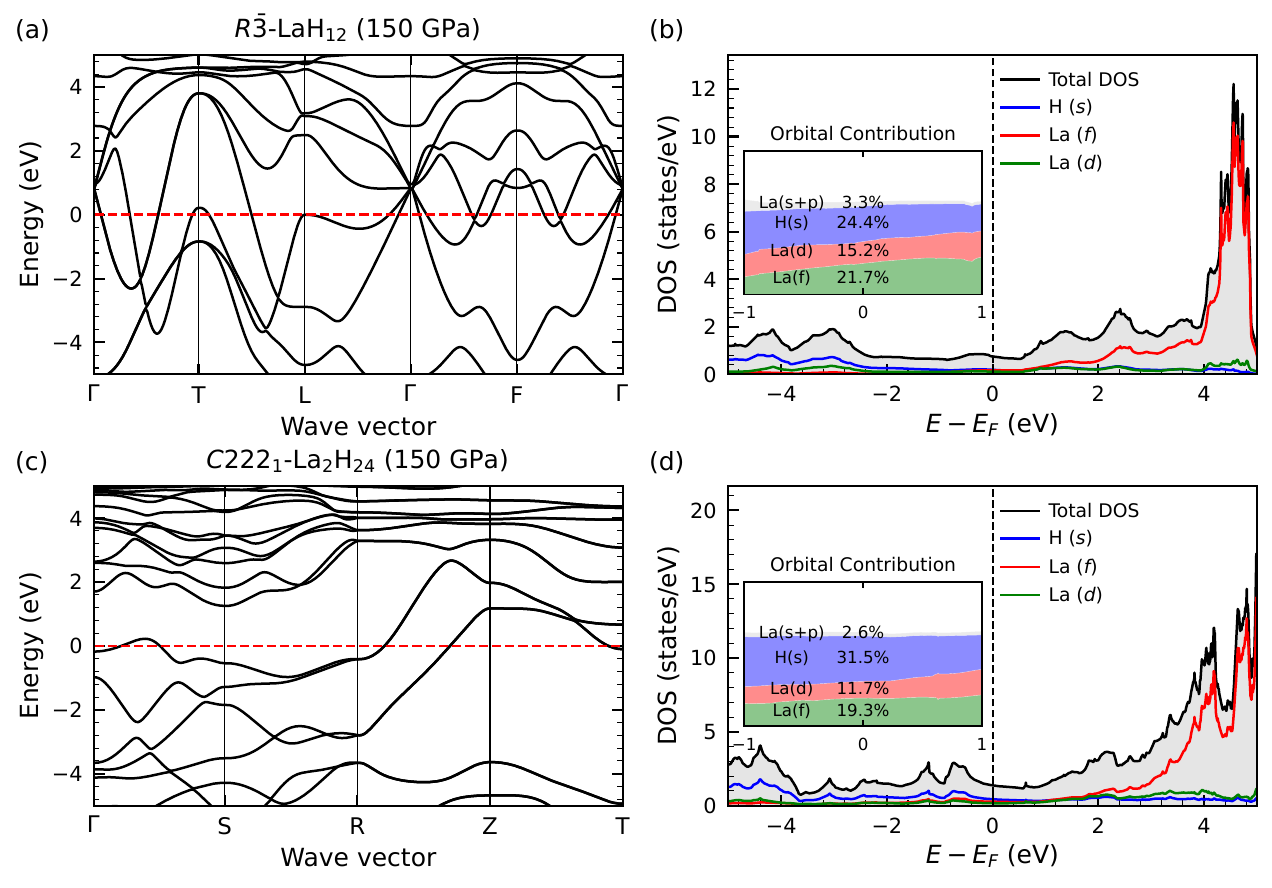}
\caption{Electronic band structure, partial DOS and orbital contribution for the (a, b) $R\bar{3}-$LaH$_{12}$ and (c, d) $C222_1$-La$_2$H$_{24}$ phases at 150 GPa. All considered LaH$_{12}$ candidates are metals under pressure of interest.}\label{fig:electronic_props}
\end{figure}

For the optimal phase $R\bar{3}$-LaH$_{12}$ (see Figure~\ref{fig:electronic_props}a, b) the orbital decomposition at the Fermi level yields La($s+p$) $\approx$ 3.3\%, H($s$) $\approx$ 24.4\%, La($f$) $\approx$ 21.7\%, and La($d$) $\approx$ 15.2\%. The computationally obtained metallicity distinguishes our candidates from the very first experimental synthesis of lanthanum dodecahydrides, where a sample of higher hydrogen content (assigned to LaH$_{12}$) was described as an insulator \cite{bib1}. The predicted metallic character and La–H hybridization are consistent with the possible experimental observation of the onset of superconductivity in LaH$_{12}$ samples \cite{bib2}. Contactless radio-frequency (RF) transmission measurements indicate a $T_c$(onset) up to 280 K at 165 GPa in the magnetic field \cite{bib2}.

However, attributing this high $T_c$ exclusively to the derived phase $R\bar{3}$-LaH$_{12}$ is problematic, as the survival of H$_2$ molecules localizes electrons, decreasing the superconducting temperature. High-pressure polyhydride synthesis yields multiphase mixtures, strongly dependent on local kinetic pathways, thermodynamic conditions and, indeed, experimental transport studies explicitly note that such samples are often mixed phases containing multiple components \cite{bib31}. Driven by the experimental detection of a hexagonal sublattice at 165 GPa \cite{bib2}, we assume that the measured high-$T_c$ signal may originate from the coexisting hexagonal phase. Our electronic structure calculations for the corresponding pseudohexagonal phase $C222_1$-La$_2$H$_{24}$ confirm a metallic state (see Figure~\ref{fig:electronic_props}c, d), where the orbital decomposition at the Fermi level yields La($s+p$) $\approx$ 2.6\%, H($s$) $\approx$ 31.5\%, La($f$) $\approx$ 19.3\%, and La($d$) $\approx$ 11.7\%.

To further expand our search space, an unconstrained evolutionary algorithm using USPEX coupled with the machine-learned MatterSim potential \cite{bib4,bib6,bib8} was executed to sample the stoichiometries of LaH$_{11}$, LaH$_{12}$ and LaH$_{13}$. This global search independently identified phase $Immm$-LaH$_{11}$ and $C2/m$-LaH$_{13}$ at 150 GPa as candidates with low enthalpy (see Figures S2 in the Supplementary Information). The generation of LaH$_{12}$ structures at 150 GPa was provided with a lanthanum atom range of 1 to 4 in the unit cell that led to the phase $Immm$-LaH$_{12}$ (see Figures S3 in the Supplementary Information), while successfully recovering previously known structural end-members. However, none of these structures correspond to experimentally observed XRD patterns. Furthermore, previously reported high-symmetry structure $Pm\bar{3}$-ScH$_{12}$ \cite{bib4} was used for direct chemical substitution (Sc $\rightarrow$ La), transforming this configuration into a phase $Pm\bar{3}$-LaH$_{12}$ at 150 GPa, to obtain another independent candidate for our comparative analysis. Electronic band structures and total/projected DOS confirm that the identified candidates $Pm\bar{3}$-LaH$_{12}$ and $Immm$-LaH$_{12}$  are metallic at 150 GPa (see Figures S4 in the Supplementary Information), with multiple dispersive bands crossing the Fermi level. 

\begin{figure}[!htbp]
    \centering
    \includegraphics[width=0.98\columnwidth]{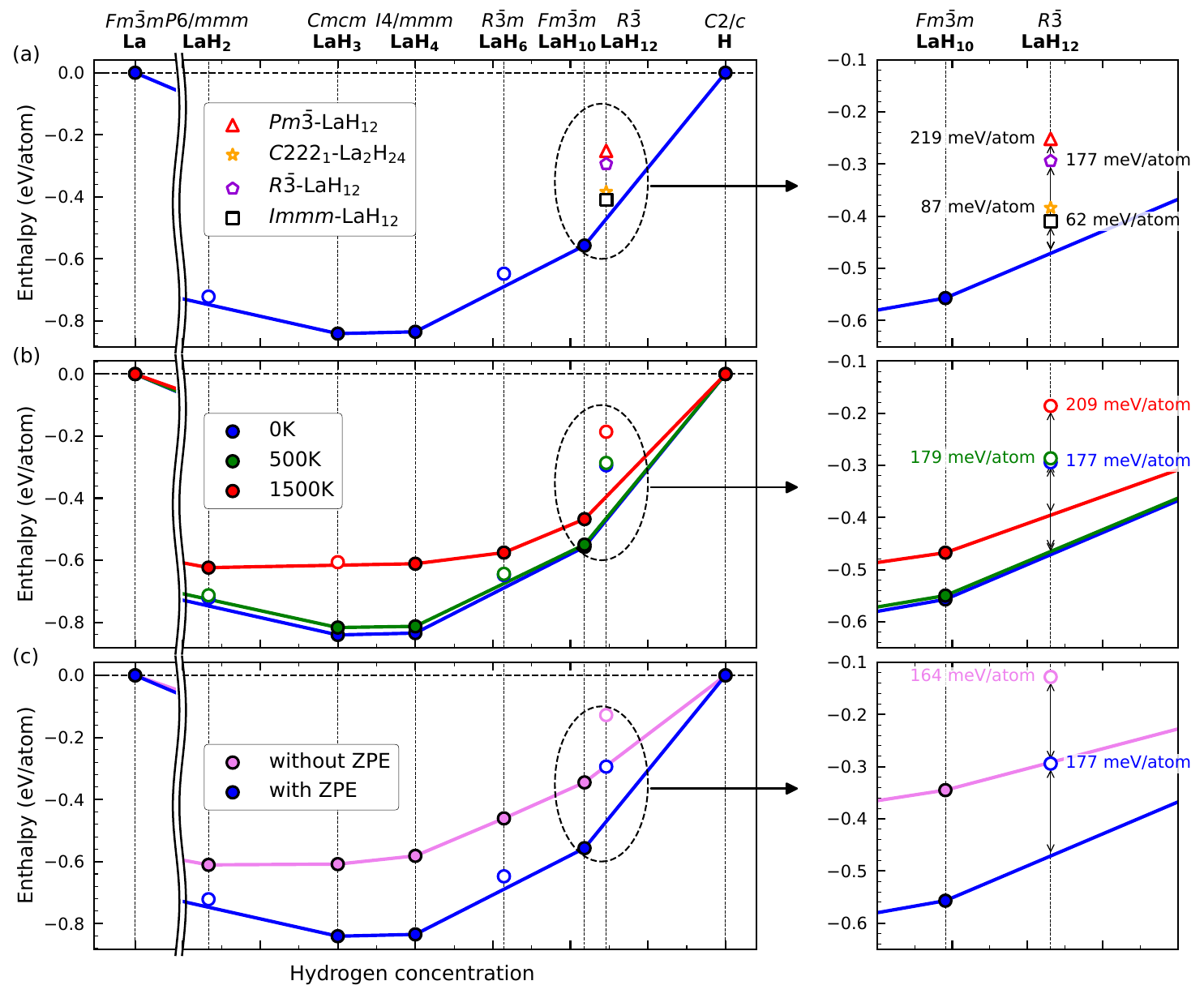}
    \caption{Thermodynamic convex hulls of the La-H system at 150 GPa. (a) Convex hull (with ZPE) showing the formation enthalpies per atom. The zoomed-in right panel highlights the precise energy distances above the convex hull for the key metastable phases of LaH$_{12}$: $Pm\bar{3}$ $C222_1$, $R\bar{3}$, and $Immm$. (b) Temperature dependence of the convex hull at 0~K (blue), 500~K (green), and 1500~K (red) highlighted for phase $R\bar{3}$ on the right panel. (c) Effect of ZPE corrections on the thermodynamic stability at 0~K highlighted for phase $R\bar{3}$ on the right panel.
    }
    \label{fig:convex_hull}
\end{figure}

To evaluate thermodynamic stability, we constructed the formation enthalpy convex hull over the entire La–H system (incorporating $Fm\bar{3}m$-La, $P_6/mmm$-LaH$_2$, $Cmcm$-LaH$_3$, $I4/mmm$-LaH$_4$, $R\bar{3}m$-LaH$_6$, $Fm\bar{3}m$-LaH$_{10}$ and $C2/c$-H) at 150 GPa with ZPE corrections (see Figure~\ref{fig:convex_hull}a). It should be noted that the thermodynamic instability of the LaH$_{12}$ stoichiometry relative to the LaH$_{10}$–H tie-line has been previously reported in early theoretical investigations \cite{bib32}. The high-symmetry phase $Pm\bar{3}$-LaH$_{12}$ with H$_2$ pseudomolecules lies $\approx$~219~meV/atom above the convex hull, and the phase $Immm$-LaH$_{12}$, found by fixcomp USPEX search, lies $\approx$~62~meV/atom above the convex hull. Candidates $R\bar{3}$-LaH$_{12}$ and $C222_1$-La$_2$H$_{24}$ derived from the experimental lattice are located at $\approx$~177~meV/atom ($\approx$~164~meV/atom without ZPE, as shown in Figure~\ref{fig:convex_hull}c) and $\approx$~87~meV/atom above the convex hull, respectively. Furthermore, elevating the temperature does not stabilize this structure: the enthalpy distance of the phase $R\bar{3}$-LaH$_{12}$ from the convex hull increases to $\approx$~179~meV/atom at 500~K and to $\approx$~209~meV/atom at 1500~K (Figure~\ref{fig:convex_hull}b).
We propose that bridging this thermodynamic gap requires moving beyond simple idealized unit cells with Z = 1, 2 to supercells with larger Z. The synthesized material may be stabilized by factors currently omitted from standard optimizations, such as exploring substantially larger supercells (e.g., LaH$_{12}\rightarrow$ La$_2$H$_{24} \rightarrow$ La$_3$H$_{36}$) to capture degenerate hydrogen orderings, non-stoichiometric compositions (e.g., La$_3$H$_{35}$ to La$_3$H$_{37}$), extended defects in the hydrogen sublattice, or even extrinsic stabilization driven by light-element impurities (Ga, C, N, O, ...) which present in high pressure chamber. However, such an expansive compositional search increases the configurational degrees of freedom, rendering it beyond the scope of this study. Nevertheless, this deviation underscores that our theoretical $R\bar{3}$ and $C222_1$ phases serve as a first-order approximations of the complex experimental structure. Our work shows that found LaH$_{12}$ structures are dynamically stable metals, have XRD patterns and unit cell parameters close to experimental ones, and are superconductors. The relatively low $T_c$s and thermodynamic stability are drawbacks of the proposed candidates.

\begin{figure}[!htbp]
\centering
\includegraphics[width=0.98\columnwidth]{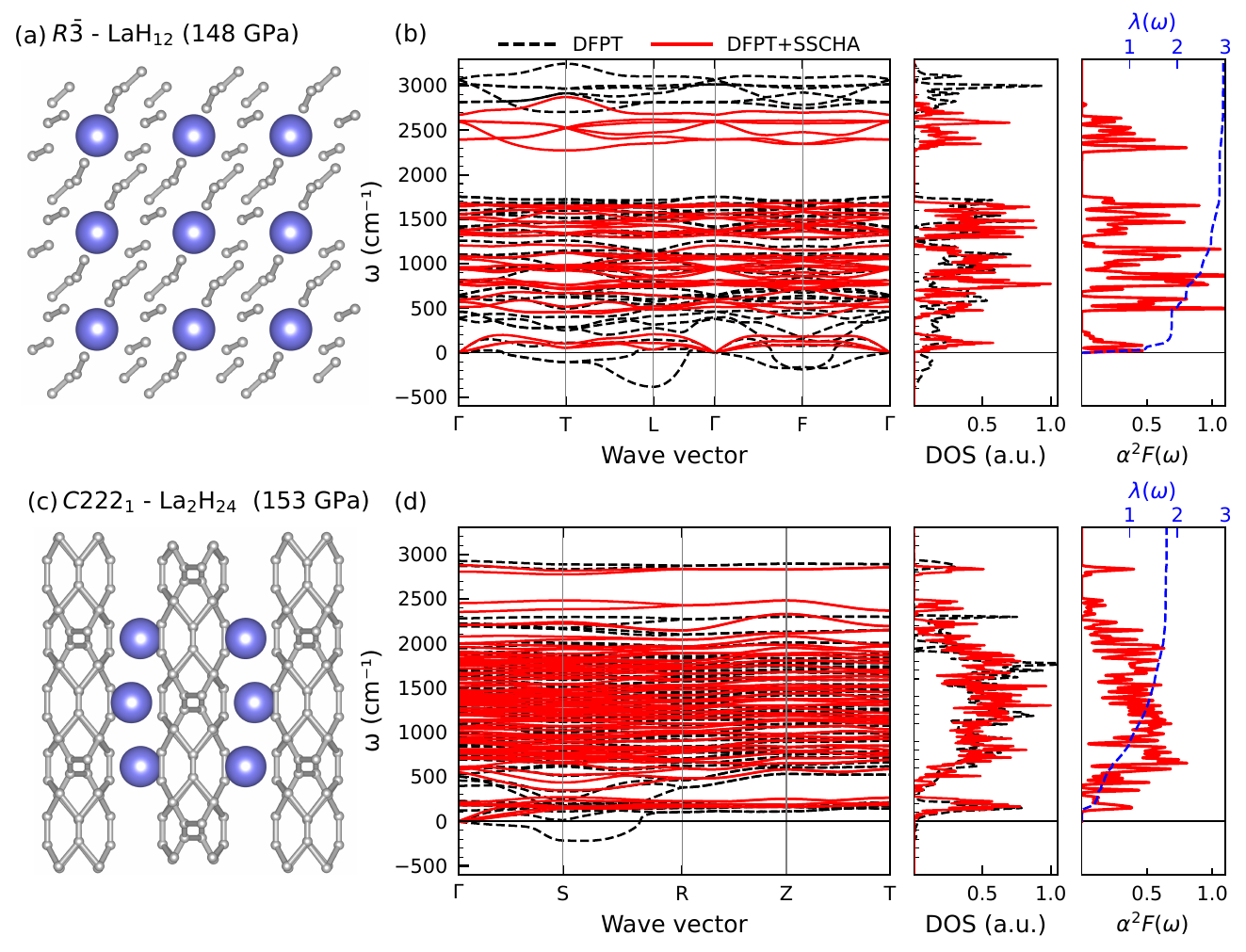}
\caption{Phonon band structures and superconducting properties of LaH$_{12}$. (a, c) Crystal structures of the $R\bar{3}$-LaH$_{12}$ (148 GPa) and $C222_1$-La$_2$H$_{24}$ (153 GPa) metastable phases, respectively. Anharmonic pressures differ from the initial harmonic pressures. (b, d) Phonon dispersions and DOS, where black dashed lines represent the harmonic approximation (DFPT) and red solid lines denote the anharmonic phonon spectra obtained via the SSCHA method. Eliashberg spectral function $\alpha^2F(\omega)$ and the integrated electron-phonon coupling constant $\lambda(\omega)$ are represented by red solid lines and blue dashed lines, respectively, for both phases.}\label{fig:sscha_props}
\end{figure}

The standard harmonic approximation fails to describe the lattice dynamics of hydrogen-dense polyhydrides due to the quantum effects of hydrogen atoms. Our calculations reveal imaginary modes in the harmonic phonon dispersions for LaH$_{12}$ (over symmetry groups of $Pm\bar{3}$, $C222_1$, $R\bar{3}$, and $Immm$ shown in Figures S5 of the Supplementary Information), concentrated in the low-frequency spectral region dominated by La displacements and hydrogen translational motions. This behavior mirrors the well-documented physics of $fcc$-LaH$_{10}$, where harmonic calculations predict structural distortion \cite{bib17}, but the inclusion of quantum effects reveals a high-symmetry structure that has become the true ground state \cite{bib17}. For these reasons, we take into account quantum fluctuations by applying the SSCHA method to evaluate the dynamical stability of the candidates obtained (Figures~\ref{fig:sscha_props}a, c), which optimally reproduce the experimentally observed phases \cite{bib2}. The calculated phonon dispersion relations and the corresponding phonon DOS are presented in Figures~\ref{fig:sscha_props}b, d. While the harmonic approximation exhibits dynamical instabilities, the anharmonic phonon spectra derived from the SSCHA method demonstrate hardening of the acoustic and optical branches, eliminating all imaginary modes. This confirms that quantum fluctuations dynamically stabilize both $R\bar{3}$-LaH$_{12}$ and $C222_1$-La$_2$H$_{24}$ phases at the pressure range 100--200 GPa (see Figure S7 and Table S1 in the Supplementary Information). Notably, the potential energy favors the distorted structure of the phase $R\bar{3}$-LaH$_{12}$ by 15 meV/atom (see Figures S1 b, c, e, f in the Supplementary Information), but the inclusion of anharmonicity stabilizes the distorted structure (see Figure S6 in the Supplementary Information) with the close difference of Helmholtz free energy 1.4 meV/atom lower compared to its undistorted counterpart. 

Furthermore, we evaluated both Eliashberg spectral functions $\alpha^2F(\omega)$ (Figures~\ref{fig:sscha_props}b, d) and the corresponding EPC parameters, ensuring strict convergence with respect to the electronic smearing parameter (see Figures S9 and S10 in the Supplementary Information). The calculated EPC constant ($\lambda$), logarithmic average frequency ($\omega_{\log}$), and estimated superconducting transition temperature ($T_c$) are summarized in Table~\ref{tab:epc_params}. The primary contribution to the EPC in the phase $R\bar{3}$-LaH$_{12}$ originates from low-frequency vibrations associated with the heavy La atoms. As noted previously, the presence of H$_2$ molecules suppresses the overall superconducting properties, yielding very low $\omega_{\log} = 11$~meV, $\lambda \approx 2.91$, and formal $T_c$(AD)~$\approx 30$~K. In contrast, the phase $C222_1$-La$_2$H$_{24}$ exhibits a robust EPC (with $\lambda \approx 1.83$ and $\omega_{\log} = 87.5$~meV), driven by contributions from mid- to high-frequency modes of the hydrogen clathrate network, resulting in an elevated $T_c \approx 150$~K. Using recent empirical analyses that established a theoretical upper bound $T_c(A_1)$ for high-pressure hydrides \cite{bib33}, we computed this integral parameter $A_1$ for our structures. The phase $R\bar{3}$ yields a substantial $A_1 = 680$~THz$^2$, where $T_c(A_1)$ is theoretically bounded up to $\approx 125$~K, while the phase $C222_1$ yields $A_1 = 886$~THz$^2$, approaching closely the optimal boundary of $T_c(A_1) \approx 170$~K near position of H$_3$S on the $T_c$ vs $A_1$ plot.

\begin{table}[!htbp] 
\caption{The EPC parameters of the studied phase modifications of LaH$_{12}$. The table summarizes the electron-phonon coupling parameter ($\lambda$), logarithmically averaged frequency ($\omega_{\log}$), integral first-order moment of $\alpha^2F(\omega)(A_1)$, the Allen-Dynes superconducting transition temperature ($T_c$), and the theoretical upper bound $T_c(A_1)$.}
\label{tab:epc_params}
\centering
\begin{tabularx}{\columnwidth}{lCCCCC}
\toprule
\multirow{2}{*}{Structure} & \multirow{2}{*}{$\lambda$} & $\omega_{\log}$ & $A_1$ & $T_c$(AD) & $T_c(A_1)$ \\
 & & (meV) & (THz$^2$) & (K) & (K) \\
\midrule
$R\bar{3}$-LaH$_{12}$ & 2.91 & 11.0 & 680 &  30 & $\approx 125$ \\
$C222_1$-La$_2$H$_{24}$ & 1.83 & 87.5 & 886 & 150 & $\approx 170$ \\
\bottomrule
\end{tabularx}
\end{table}


Returning to the case of $R\bar{3}$-LaH$_{12}$ at 150 GPa and its unrealistically soft phonon modes, we were interested in investigating what would happen if these soft phonon modes were suppressed or removed using a suitable smoothing function that preserves the form $\alpha^2F(\omega) \propto \omega^2$ in the vicinity of zero phonon energy (for example, formula S1 in Supplementary Information). Simple estimates show that suppression the soft phonon modes greatly increase $\omega_{\log}$ and $T_c$ of $R\bar{3}$-LaH$_{12}$ to nearly 100 K, in agreement with estimates based on $A_1$ (Supplementary Figure S12).

\begin{figure}[!htbp]
\centering
\includegraphics[width=0.98\columnwidth]{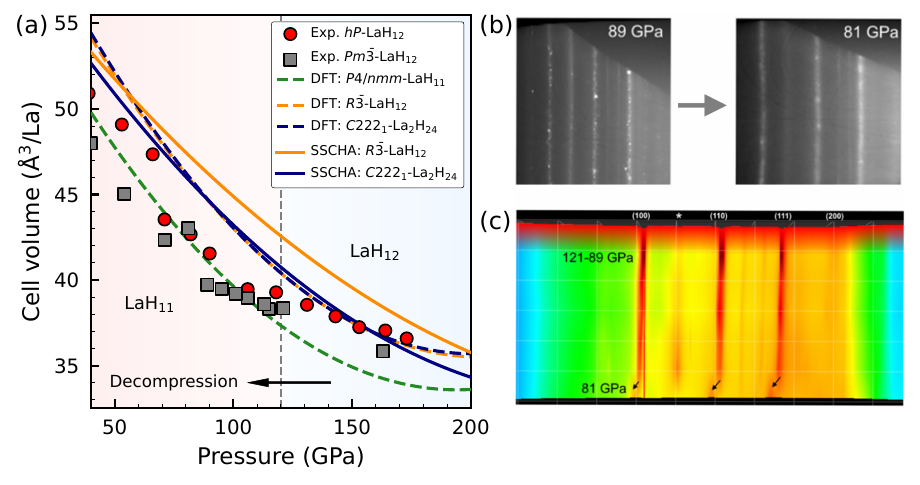}
\caption{Equations of state for the LaH$_{11}$ and LaH$_{12}$ phases. The solid orange line represents the theoretical EoS for the $R\bar{3}$-LaH$_{12}$ phase, while the blue line is for the $C222_1$-La$_2$H$_{24}$ phase at 150 GPa. Experimental data points obtained from Ref. \cite{bib2} during decompression run are denoted by red circles and grey squares. (b) A qualitative change in the diffraction patterns of a mixture of $Pm\bar{3}$ and $hP$-LaH$_{12}$ during decompression in the range from 89 to 81 GPa. A significant broadening of the diffraction peaks can be observed, caused by the fracturing of sample’s crystallites due to a possible phase transition. (c) Evolution of LaH$_{12}$ diffraction patterns during decompression in the form of an intensity heatmap. A sharp change in the position and width of the diffraction peaks in the 89–81 GPa range, caused by a sharp increase in the unit cell volume.
}
\label{fig:eos}
\end{figure}

To further investigate the structural nature of synthesized polyhydrides, we compared theoretical anharmonic equations of state (EoS) with experimental data obtained during decompression cycles in Ref. \cite{bib2} (Figures~\ref{fig:eos}). It is evident that the high-symmetry phase $R\bar{3}$-LaH$_{12}$ overestimates the unit cell volume per La atom over the entire pressure range. This suggests that this low-density phase is unlikely to be synthesized at high pressure. However, the phase $C222_1$-La$_2$H$_{24}$ exhibits a more dense structural packing and brings the theoretical prediction closer to the experimental observations. Furthermore, the experimental EoS exhibits a deviation from the predicted EoS as the pressure decreases below $\approx$~120~GPa (indicated by the dashed vertical line in Figures~\ref{fig:eos}), where the experimental cell volumes deviate even further from the extrapolated LaH$_{12}$ trend. This volume decrease relative to that expected from the equation of state is a signature of pressure-induced continuous hydrogen loss that leads to a transition from the stoichiometry of LaH$_{12}$ to a hydrogen-deficient phase LaH$_{11}$.

Indeed, in the experiment with LaH$_{12\pm1}$ in Ref. \cite{bib2}, we observed a similar sharp structural transition and change in the character of the diffraction rings, but at a lower pressure: between 89 and 81 GPa (Figures 5b, c). In this range, the narrow diffraction rings and point diffraction reflections from micron- and submicron-sized crystals of the $Pm\bar{3}$ phase broaden and acquire a diffuse character (Figure 5b). The same is observed for the diffraction bands of the hexagonal phase (* in Figure 5c). At the same time, a sharp increase in the unit cell volume is observed (Figure 5c), which may indicate a rearrangement of the hydrogen sublattice in this pressure range.

\section{Conclusion}\label{sec4}

In summary, we have addressed the problem of resolving the hydrogen sublattice in LaH$_{12}$ phases under megabar pressures. The implementation of the EnvXGen algorithm and similarity metric enabled the search of $R\bar{3}$-LaH$_{12}$ and $C222_1$-La$_2$H$_{24}$ phases that are in acceptable agreement with the experimental diffraction patterns. The generated structures are thermodynamically metastable metals that lie above the La-H convex hull, but their calculated equations of state capture both the pressure-dependent unit-cell volumes and the possible LaH$_{12} \rightarrow$ LaH$_{11}$ evolution due to hydrogen loss, observed experimentally during the decompression of LaH$_{12}$. Quantum zero-point fluctuations dynamically stabilize both the $R\bar{3}$-LaH$_{12}$ and $C222_1$-La$_2$H$_{24}$ phases while preserving the structural match with the experimental XRD profiles.

$R\bar{3}$-LaH$_{12}$ structure hosts a highly elevated electron-phonon coupling constant ($\lambda \approx 2.91$), driven by soft phonons that suppress the effective logarithmic frequency ($\omega_{\log} = 11$~meV), limiting its critical temperature to $T_c$(AD)$~\approx30$~K. Suppressing these soft phonon modes ca cause $T_c$ to rise up to $\sim100$K. In contrast, $C222_1$-La$_2$H$_{24}$ features a dense atomic-like hydrogen clathrate network that maintains a strong coupling ($\lambda \approx 1.83$), provides high-frequency hydrogen vibrations ($\omega_{\log} = 87.5$~meV) and results in higher $T_c \approx 150$~K. It is expected that increasing the size of the supercell (Z > 2) to identify the most favorable hydrogen sublattices, including hydrogen defects and non-stoichiometric compositions, and accounting for the influence of impurities, could significantly improve the agreement between theoretical calculations and experimental data.

\section*{Acknowledgements}
D.V.S. and D.Z. would like to express their gratitude to the National Natural Science Foundation of China (NSFC), Research Fund for International Excellent Young Scientists, grant number W2632008, for the financial support. D.V.S., D.Z., V.V.S. thank the NSFC International Collaboration Fund for Creative Research Teams, grant No. W2641002, for financial support. This work was supported by the National Key Research and Development Program of China (grant 2023YFA1608900, subproject 2023YFA1608903). V.V.S. acknowledges financial support from the Shanghai Science and Technology Committee, China (No. 22JC1410300) and Shanghai Key Laboratory of Materials Frontier Research in Extreme Environments (MFree), China (No. 22dz2260800).

\section*{Author contributions}
K. A.: conceptualization; methodology; investigation; formal analysis; writing – original draft; D.V.S.: conceptualization; methodology; investigation; formal analysis; writing; conceptualization; methodology; investigation; formal analysis; writing; supervision. D.Z.: investigation; funding acquisition. V.V.S.: conceptualization; supervision; funding acquisition; writing - review and editing. A.G.K.: supervision; funding acquisition; writing - review and editing. All authors discussed the results and approved the final manuscript.

\section*{Data availability}
The authors declare that the main data supporting findings of this study are contained within the paper and the Supporting Information. Data are available from the corresponding author upon reasonable request.

\section*{Competing interests}
The authors declare no conflicts of interest. 

\bibliographystyle{elsarticle-harv}
\bibliography{bibliography}

\end{document}